\documentclass[prd, amsfonts, twocolumn, nofootinbib, showpacs]{revtex4-2}
\usepackage{graphicx, epsfig}
\usepackage{color}
\usepackage{amsmath}
\usepackage{amssymb}
\usepackage{physics}
\newcommand{\be}{\begin{equation}}
\newcommand{\ee}{\end{equation}}
\newcommand{\bea}{\begin{eqnarray}}
\newcommand{\eea}{\end{eqnarray}}

\newcommand{\gapp}{\mathrel{\raise.3ex\hbox{$>$}\mkern-14mu
\lower0.6ex\hbox{$\sim$}}}
\newcommand{\lapp}{\mathrel{\raise.3ex\hbox{$<$}\mkern-14mu
\lower0.6ex\hbox{$\sim$}}}
\def\bbox{{\,\lower0.9pt\vbox{\hrule \hbox{\vrule height 0.2 cm
\hskip 0.2 cm \vrule  height 0.2 cm}\hrule}\,}}

\begin{document}
\title{Superluminal propagation along the brane in space with extra dimensions  }
\author{De-Chang Dai$^{1,2}$\footnote{communicating author: De-Chang Dai,\\ email: diedachung@gmail.com\label{fnlabel}}, Dejan Stojkovic$^3$}
\affiliation{ $^1$ Department of Physics, national Dong Hwa University, Hualien, Taiwan, Republic of China}
\affiliation{ $^2$ CERCA, Department of Physics, Case Western Reserve University, Cleveland OH 44106-7079}
\affiliation{ $^3$ HEPCOS, Department of Physics, SUNY at Buffalo, Buffalo, NY 14260-1500}

\begin{abstract}
\widetext
We demonstrate that a model with extra dimensions formulated in \cite{Csaki:1999mp}, which fatefully reproduces Friedmann-Robertson-Walker (FRW) equations on the brane, allows for an apparent superluminal propagation of massless signals. Namely, a massive brane curves the spacetime and affects the trajectory of a signal in a way that allows a signal sent from the brane through the bulk to arrive (upon returning) to a distant point on the brane faster than the light can propagate along the brane. In particular, the signal sent along the brane suffers a greater gravitational time delay than the bulk signal due to the presence of matter on the brane. While the bulk signal never moves with the speed greater than the speed of light in its own locality, this effect still enables one to send signals faster than light from the brane observer's perspective. For example, this effect might be used to resolve the cosmological horizon problem. In addition, one of the striking observational signatures  would be arrival of the same gravitational wave signal at two different times, where the first signals arrives before its electromagnetic counterpart.       
We used GW170104 gravitational wave event to impose a strong limit on the model with extra dimensions in question.   
\end{abstract}


\maketitle

\section{Introduction}

Superluminal propagation of a signal in some theoretical model is usually associated with problems, most notably causality. Since not so many physicists are willing to sacrifice causality (at least not at the macroscopic level), there is no vast literature on this topic. 

\begin{figure}[h]
\includegraphics[width=4cm]{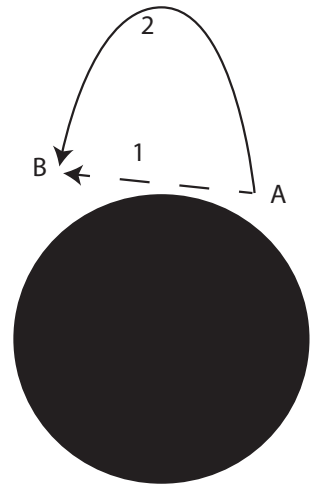}
\caption{The spacetime is highly curved nearby a compact object like a black hole. If two light signals, 1 and 2, are sent from A simultaneously, under the right conditions, the signal 2 can arrive to B before the signal 1. While the speed of light in its own locality always remains the same, it appears to the observers located at A and B that the signal 2 traveled faster than the light signal 1.       
}
\label{gravity}
\end{figure}

If we avoid propagation of a signal with intrinsically superluminal velocities, we are not left with many options. A light signal cannot overtake itself in its own locality by definition. However, in a curved space, one can easily imagine a situation where a light signal sent along a certain trajectory can overtake another light signal sent along a different trajectory. 
Black holes are templates for interesting phenomena in curved space. Any signal propagating in a vicinity of a black hole will suffer a significant redshift (or equivalently gravitational time delay). In an extreme case, light emitted exactly from the horizon will be practically stopped. Imagine a situation like in Fig. \ref{gravity} where one signal (labeled by 1) travels very close to the black hole horizon from the point A to the point B, while another signal (labeled by 2) travels also from A to B, first away from the black hole and then back. If these two signals are sent from A simultaneously, under the right conditions, the signal 2 can arrive to B before the signal 1.

\begin{figure}[h]
\includegraphics[width=8cm]{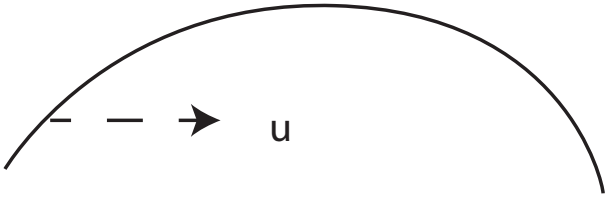}
\caption{ A signal traveling through the bulk can overtake the light signal traveling at the speed of light along the curved brane. Thus, an observer confined on the brane might register an apparently superluminal propagation of a signal.  
}
\label{curved-brane}
\end{figure}

A new playground was introduced in the context of the brane world models \cite{Arkani-Hamed:1998jmv,Antoniadis:1998ig,Randall:1999ee,Randall:1999vf,Starkman:2001xu,Starkman:2000dy}
where all the standard model particles are located on a subspace (brane) in a higher dimensional universe. Gravity is allowed to propagate everywhere including the bulk. In such a setup, it is easy to construct a shortcut  if the brane is curved. For example, Fig. \ref{curved-brane} shows that a signal traveling through the bulk can overtake the light signal traveling at the speed of light along the curved brane. Thus, an observer confined on the brane might register an apparently superluminal propagation of a signal.  In \cite{Chung:1999xg}, it has been proposed that such shortcuts can be used to solve the cosmological horizon problem. Similar cases were studied in \cite{Abdalla:2002ir,Abdalla:2004ica}. Obviously, these shortcuts are not a generic feature of all the brane world models, and they require elaborate setups.  

In an interesting work in \cite{Greene:2022uyf,Greene:2022urm,Polychronakos:2022uam}, it was shown that an observer on a moving brane with compact extra dimensions can also register an apparently superluminal propagation of a signal.  

\section{Model}

In this paper we extend the previous work to fix some of the shortcomings of the existing models.   We consider the concrete model formulated by Csaki et. al. in \cite{Csaki:1999mp}. 
The motivation for this model comes from the appearance of non-conventional late time cosmologies in Randall-Sundrum models where the radion field (which dictates the distance between the two branes) is not stabilized \cite{Brax:2003fv,Kanti:2002zr,Barger:2000wj}. Such problems disappear in the presence of a stablizing potential, and the ordinary FRW (Friedmann-Robertson-Walker) equations are reproduced, with the expansion driven by the sum of the physical values of the energy densities on the two branes and in the bulk.

\begin{figure}[h]
\includegraphics[width=8cm]{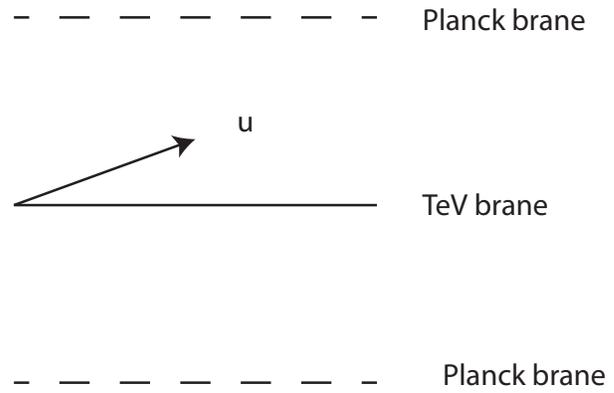}
\caption{ An extra dimension is compactified on a $S^1/Z_2$ manifold.  The bulk is mirror symmetric around the brane where the standard model particles are located (the TeV brane). A massless signal emitted from the TeV brane can travel through the bulk, reach the upper Planck brane which is identified with the lower Planck brane, and then return to the original TeV brane.    
}
\label{2-brane}
\end{figure}

 In contrast with \cite{Greene:2022uyf}, the branes are fixed in the bulk and are not moving. The extra dimension is compactified on a  $S^1/Z_2$ manifold (as in Fig. \ref{2-brane}). Thus, unlike \cite{Chung:1999xg}, a signal sent into the bulk is guaranteed to return to the brane. The metric is written as  
\begin{equation}
d\tau^2=n(y,t)^2 dt^2 -a(y,t)^2 (dx_1^2+dx_2^2+dx_3^2)-b(y,t)^2 dy^2 .
\end{equation}

and the components of the Einstein tensor are 
\begin{eqnarray}
G_{00}&=&3\Bigg( \Big( \frac{\dot{a}}{a}\Big)^2+\frac{\dot{a}\dot{b}}{ab}-\frac{n^2}{b^2}\Big(\frac{a''}{a}+\Big(\frac{a'}{a}\Big)^2-\frac{a'b'}{ab}\Big)\Bigg)\\
G_{ii}&=&\frac{a^2}{b^2}\Bigg(\Big(\frac{a'}{a} \Big)^2+2\frac{a'n'}{an}-\frac{b'n'}{bn}-2\frac{b'a'}{ba}+2\frac{a''}{a}+\frac{n''}{n}\Bigg)+\nonumber\\
&&\frac{a^2}{n^2}\Bigg(-\Big(\frac{\dot{a}}{a}\Big)^2+2\frac{\dot{a}\dot{n}}{an}-2\frac{\ddot{a}}{a}-\frac{\dot{b}}{b}\Big(2\frac{\dot{a}}{a}-\frac{\dot{n}}{n}\Big)-\frac{\ddot{b}}{b}\Bigg)\\
G_{05}&=&3\Bigg(\frac{n'\dot{a}}{na}+\frac{a'\dot{b}}{ab}-\frac{\dot{a}'}{a}\Bigg)\\
G_{55}&=&3\Bigg(\frac{a'}{a}\Big(\frac{a'}{a} +\frac{n'}{n}\Big)-\frac{b^2}{n^2}\Big(\frac{\dot{a}}{a}\Big(\frac{\dot{a}}{a}-\frac{\dot{n}}{n}\Big)+\frac{\ddot{a}}{a}\Bigg)
\end{eqnarray}
In this setup there are two branes, one is located at $y=0$ and another at $y=1/2$. The brane at $y=0$ is called the "Planck brane", while the brane at $y=1/2$ where all the standard model particles are localized is called the "TeV brane". So we assume that our universe is at the TeV brane. 

The energy momentum tensor for this configuration is approximately 
\begin{eqnarray}
T^b_a &=&\frac{\delta(y)}{b} diag (\rho_*,-p_*,-p_*,-p_*,0)+\nonumber\\
 &&\frac{\delta(y-\frac{1}{2})}{b} diag (\rho,-p,-p,-p,0) ,
\end{eqnarray}
where $p$ and $p_*$ denote pressure on the TeV and Planck branes respectively, while $\rho$ and $\rho_*$ are their corresponding energy densities. The space time is $S^1/Z_2$ symmetric, i.e. it spans from $y=0$ to $y=1$ and is mirror symmetric at $y=0$ and $y=1/2$. Because of the mirror symmetry, each brane has its own mirror images. Thus, taking images into account, the locations of the TeV brane are $y=...-1/2, 1/2, 3/2, 5/2,...$, while the locations of the Planck brane are $y= ...-1,0,1,2,3,...$.

The Einstein equations are 
\begin{equation}
G_{\alpha\beta} =\kappa^2 T_{\alpha\beta} ,
\end{equation}
where $\kappa^2=1/2M^3$, while $M$ is the five dimensional Planck scale. Apart from matter on the branes, a radion field with an appropriate potential is introduced to stabilize the extra dimension. Here we are not going into unnecessary details and quote an approximate solution from the appendix B in \cite{Csaki:1999mp}: 

\begin{eqnarray}
a&=&a_0(t) (1+\alpha \rho_*(t) (y-\frac{1}{2})^2 +\beta \rho(t)y^2)\\
n&=& (1+\gamma \rho_*(t) (y-\frac{1}{2})^2 +\lambda \rho(t)y^2)\\
b&=&b_0(1+\delta b) .
\end{eqnarray}

From the jump conditions we get 
\begin{eqnarray}
\alpha&=&\beta=\frac{\kappa^2 b_0}{6}\\
\gamma&=&-\frac{(2+3\omega_*)\kappa^2 b_0}{6}\\
\lambda&=&-\frac{(2+3\omega)\kappa^2 b_0}{6}\ ,
\end{eqnarray}
where $p = \omega \rho$ and $p_* = \omega_* \rho_*$, while $\delta b=O(\rho^2,\rho\rho_*, \rho_*^2)$. This solution is valid between  $y =0$ and $y=1/2$. For the other regions, the solutions are obtained by reflecting around $y = 0$ or $y =1/2$. The Einstein equations are satisfied to the first order in $\kappa^2b_0 \rho$. The scale factor $a_0$ is obtained from the Friedmann equations
\begin{eqnarray}
&&\Big(\frac{\dot{a}_0}{a_0}\Big)^2=\frac{\kappa^2}{3b_0}(\rho+ \rho_*),\\
&&\Big(\frac{\dot{a}_0}{a_0}\Big)^2 + 2 \frac{\ddot{a}_0}{a_0}= -\frac{\kappa^2}{b_0}(\omega \rho+ \omega_*\rho_*) ,
\end{eqnarray}
where, $M_p^2= b_0/\kappa^2$. 

\section{Propagation of the bulk and brane signals}

We consider signals delivered by massless particles. Since proper time has no meaning for a massless particle, we choose the coordinate time ($t$) which defines a coordinate velocity as
\begin{equation}
u^\alpha =\frac{d x^\alpha}{dt} .
\end{equation}
In this case, $u^t=1$, and the geodesic equation is  
\begin{equation}
\label{geodesic}
\frac{d^2 x^\lambda}{dt^2}=-\Gamma^\lambda_{\nu\alpha} \frac{dx^\nu}{dt}\frac{dx^\alpha}{dt} +\Gamma^t_{\nu\alpha} \frac{dx^\nu}{dt}\frac{dx^\alpha}{dt}\frac{dx^\lambda}{dt} .
\end{equation}

We assume that matter is concentrated on the TeV brane, while $\rho_*=0$, and $\omega=\omega_*=0$ on the Planck brane.  Since in this model the expansion is driven by the sum of the physical values of the energy densities on the two branes and in the bulk, we have freedom to make such choice without affecting FRW phenomenology on our brane. 
The energy density on the Tev brane is  
\begin{equation}
\rho=\frac{\rho_0}{a^3}\approx \frac{\rho_0}{a_0^3} .
\end{equation} 
Here we kept only the leading order in $a$, while $\rho_0$ is the initial density at $t=0$. The scale factor takes the form
\begin{equation}
a_0= \Big( 1+\sqrt{\frac{3\kappa^2 \rho_0}{4b_0}} t\Big)^{2/3} .
\end{equation}

We now analyze propagation of signals in this setup. A massless particle is emitted from the TeV brane with a small initial velocity component in y-direction, i.e.  $u^y_0 \neq 0$. The corresponding velocity component $u^x_0$ is obtained from $a^2(u^x_0)^2 +b^2(u^y_0)^2=n^2$, since the particle is massless. We assume that  $u^y_0 = O(\rho)$. According to Eq. \ref{geodesic}, the acceleration in $y$ direction is 
\begin{equation}
\frac{du^y}{dt} = - \Gamma^y _{\alpha\beta}u^\alpha u^\beta + \Gamma^t _{\alpha\beta}u^\alpha u^\beta u^y = O(\rho) 
\end{equation}  
Therefore, when the particle crosses the bulk, its y-direction velocity is still $O(\rho)$. 
For the component along the brane, $u^x$, the acceleration is 
\begin{equation}
\frac{du^x}{dt} = -2 \Gamma^x _{tx} u^t u^x - 2 \Gamma^x _{yx}u^y u^x  + \Gamma^t _{\alpha\beta}u^\alpha u^\beta u^x .
\end{equation} 
 The first term on the right hand side is induced by the space expansion, the second term is of order $O(\rho^2)$, while the third term is of order $O(\rho)$. If $u^y$ remains   $O(\rho)$, the deviation of $u^x$ from the initial one should be of order $O(\rho^2)$. It then is possible to construct a signal that leaves and then returns to the brane due to the $Z_2$ symmetry.

\begin{figure}[h]
\includegraphics[width=8cm]{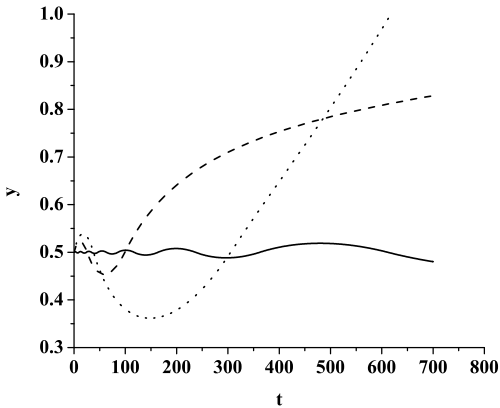}
\caption{The displacement of a massless signal in $y$-direction (bulk) sent from the TeV brane (located at $y=1/2$). We set $\kappa^2=1$, $b_0=1$ and $\rho_0=10^{-4}$. The solid, doted, and dashed lines represent the displacement along the $y$-coordinate for three initial values of velocity: $u_0^y=1\times 10^{-3}$, $u_0^y=4\times 10^{-3}$, and $u_0^y=5\times 10^{-3}$ respectively. When $u_0^y$ is small, the signal is confined nearby the brane due to gravitational attraction of matter of the brane. However, when $u_0^y$ is large enough, gravitational attraction is not able to confine the signal, and the signal can leave the TeV brane at $y=1/2$, reach the Planck brane at $y=1$ which is identified with the Planck brane at $y=0$, and come back to the original TeV brane (the return is not shown here). 
}
\label{move}
\end{figure}

Consider a signal that starts from $(x,y)=(0,1/2)$, and returns to the TeV brane at the moment T.
The displacements in x- and y-directions are  
\begin{eqnarray} \label{dis}
x(t)&=&\int_0^T u^x dt \\  
y(t)&=&\int_0^T u^y dt  \nonumber .
\end{eqnarray}
We numerically integrate Eq. (\ref{dis}) and plot the results.  Fig. \ref{move} shows the displacement of a signal in $y$-direction as a function of time. If the initial magnitude of the velocity $u_0^y$ is small, the signal cannot go far away from the brane and is just oscillating near the brane. If $u_0^y$ is large enough, it can leave the TeV brane, propagate  to the Planck brane, and return to the TeV brane due to mirror symmetry (doted line). Even in the first (oscillatory) case, the oscillation amplitude is increasing in time since the energy density is reducing due to expansion. When the density is diluted enough and gravitational attraction weakened, the signal can leave the brane in this case too.

\begin{figure}[h]
\includegraphics[width=8cm]{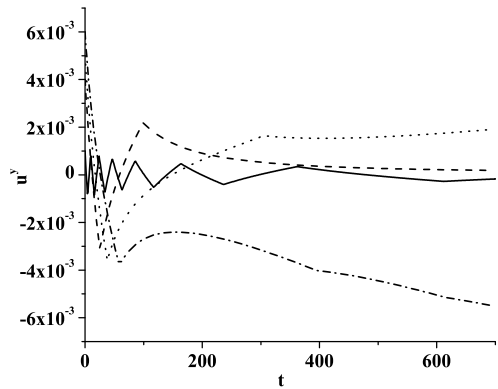}
\caption{Velocity $u^y$ as a function of time. We set  $\kappa^2=1$, $b_0=1$ and $\rho_0=10^{-4}$.  The solid, doted, dashed and dash-doted lines represent $u^y$ for four values of initial velocity: $u_0^y=1\times 10^{-3}$, $u_0^y=4\times 10^{-3}$, $u_0^y=5\times 10^{-3}$, and $u_0^y=6\times 10^{-3}$ respectively. When $u_0^y$ is small, the velocity is oscillating around $0$. However, when $u_0^y$ is large, $u^y$ increases with time. Since $u^x$ decreases due to the expansion of our universe,  $u^y$ must increase to maintain the overall speed of light.    
}
\label{uy}
\end{figure}

We plot the velocity $u^y$ as a function of time in Fig. \ref{uy}. It is clear that $u^y$ is also oscillating. Many sharp changes in velocity are noticeable since the attractive gravitational force changes direction when the signal crosses the TeV brane. The $u^x $ component  is reduced much faster than $u^y$ because our universe expands in x-direction. Therefore, the $u^y$ component increases and becomes the dominant component in velocity. In that regime, the signal moves perpendicular to the brane.     

\begin{figure}[h]
\includegraphics[width=8cm]{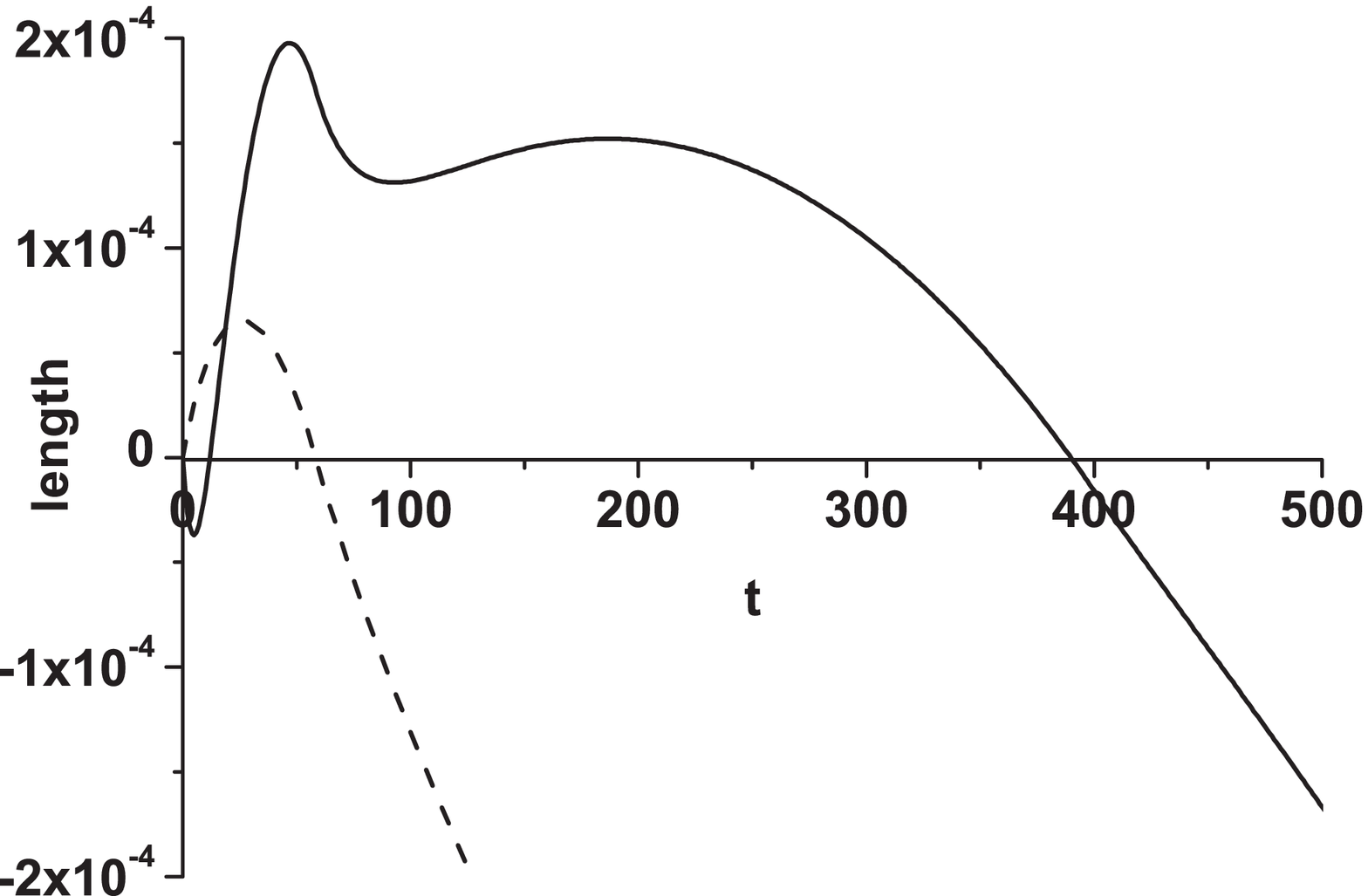}
\caption{Comparison of the signals propagating on the brane and the bulk. We set $\kappa^2=1$, $b_0=1$ and $\rho_0=10^{-4}$. We denote  the displacements in $x$ direction  of the signals that travel with $u_0^y=6\times 10^{-3}$ (bulk signal) and $u_0^y=0$ (brane signal) respectively with $x_b$ and $x_0$. The solid line tracks the difference in these displacements, i.e. $\Delta x=x_b-x_0$. The dashed line is $(y-0.5)\times 10^{-3}$. We plot it here to track the signal, and in particular to show when the signal returns to the brane.
At first, the signal on the brane moves faster than in the bulk ($\Delta x$ is negative), however at later time the bulk signal overtakes the brane signal ($\Delta x$ is positive). Upon returning to the brane, the bulk signal is ahead of the signal on the brane. Thus, an observer confined to the brane may register an apparent superluminal motion.  
}
\label{time}
\end{figure}

In Fig. \ref{time}, we  compare the propagation of the signals on the brane and the bulk. We can see that, at first, the signal on the brane moves faster than in the bulk because $u^x$ is larger than $u^y$ in magnitude. When the bulk signal propagates far enough, the situation changes and the bulk signal overtakes the brane signal. From the same figure, one can see that when the signal sent to the bulk returns to brane due to gravity, it is ahead of the signal on the brane. This clearly shows  that an observer confined to the brane may register an apparent superluminal motion.

\begin{figure}[h]
\includegraphics[width=8cm]{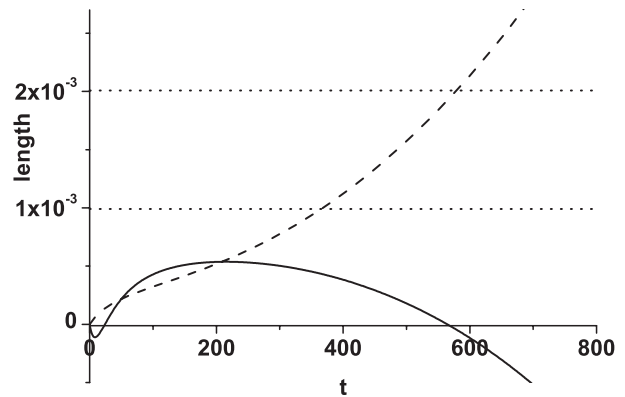}
\caption{Comparison of the signals propagating on the brane and the bulk, similar to Fig. \ref{time} but for much larger value of $u_0^y$. We set  $\kappa^2=1$, $b_0=1$ and $\rho_0=10^{-4}$.  This time, the bulk signal initial velocity is  $u_0^y=8.3\times 10^{-3}$. The solid line tracks again the difference in displacements in $x$-direction, i.e. 
$\Delta x=x_b-x_0$.The dashed line is again $(y-0.5)\times 10^{-3}$.  For convenience, we locate the TeV brane at $y=0$, while the doted lines represents two other images of the same brane. The signal can travel to the other images before the brane confined signal arrives. However, after a few round trips, this feature disappears ($\Delta x$ becomes negative) because the signal is tilted into the bulk direction.       
}
\label{time-83}
\end{figure}

For larger values of the initial bulk component of velocity  $u_0^y$, the signal is not trapped in vicinity of the brane. It is able to make a trip to the other brane and return to the original TeV brane. Fig. \ref{time-83} shows again  that a brane confined observer can observe an apparent  superluminal effect. However, this effect lasts only for a first few round-trips. Once the magnitude of $u^y$ is dramatically increased by the universe expansion, the $u^x$ component of the bulk signal is so small comparing to signal on the brane that the redshift effect cannot compensate for the velocity difference between these two signals. In this regime, the bulk signal cannot overtake the signal along the brane. 

\begin{figure}[ht!]
\includegraphics[width=8cm]{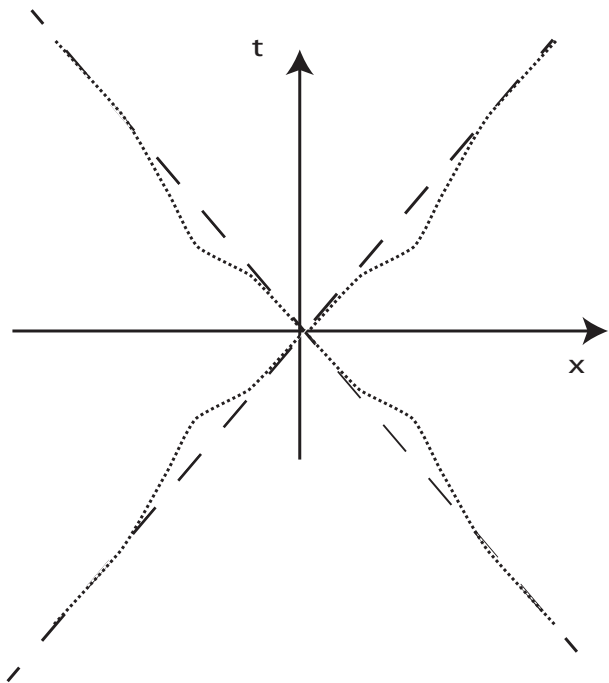}
\caption{The dashed line is the lightcone drawn (expected) by a brane confined observer. The doted line is the actual brane lightcone drawn by the bulk observer who sees the whole situation. Obviously, the expected lightcone is smaller than the actual lightcone, which gives rise of an apparent superluminality.}
\label{lightcone}
\end{figure}

In Fig. \ref{lightcone}) we plot the lightcones that nicely illustrate the whole situation. The lightcone drawn by an observer located on the brane (who can observe only signals along the brane) is smaller than light cone that includes the whole setup (the bulk and the brane). This explains the apparent superluminality observed by the brane observer. However, at late times the lightcones match again and the superluminal effect disappears.  

\section{Limits from gravitational waves}

All the discussion above was dedicated to propagation of massless signals. Therefore, practically all the conclusions will remain the same for the case of gravitational waves propagation. 
One of the striking observational signatures  would be arrival of the same gravitational wave signal at two different times, where the first signals arrives before its electromagnetic counterpart.       
In addition, echo-like signals \cite{Dong:2020odp,Abedi:2016hgu} should also be present, because there will be many signals coming from different brane images. Detailed modeling of the gravitational wave signature will be reserved for future studies.  However, we can already use some of the observed gravitational wave events to impose limits on the model in question. 

For example, the source of gravitational event GW170104 is located at luminosity distance $l=880$Mpc\cite{LIGOScientific:2017bnn}.  The time interval to see a repeated signal in the model with extra dimensions we considered here can be estimated from  
\begin{equation}
\Delta t = \frac{\sqrt{l^2+h^2} -l}{c}\approx \frac{h^2}{2lc}
\end{equation}
where $h$ is the size of the compact space in between the branes. Since the repeated signal has not been observed in 5 years, we set $\Delta t>5$years. Then it follows that $h>0.05$Mpc.
This is a pretty strong constraint on the size of extra dimensions in this model, which could not be established by any other means. We emphasize that virtually all the existing constraints on the Randall-Sundrum models and its variants are based on microscopic physics. If one is interested in solving the standard model hierarchy problem, the distance between the two branes must be microscopic. The strongest limits come from the production of the Kaluza-Klein gravitons and deviations from the Newton's inverse square law  \cite{CMS:2018ipm,Lee:2020zjt}. In that case the distance between the branes should be at most of the order of microns. However, if one abandons the solution to the hierarchy problem, then the second brane can be at a large distance from our brane. In fact, the second brane can be sent to infinity, like in the second Randall-Sundrum model \cite{Randall:1999vf}. So far there was no limit on how far can the two branes be set apart.

\section{Conclusions}

We demonstrated here that it is possible to send signals that appear superluminal from the point of view of an observer confined to a brane located in a higher dimensional universe.  
We used a variant of the Randall-Sundrum model with one extra dimension and two branes (the TeV brane that represents our universe and the Planck brane). Due to the imposed $S^1/Z_2$ compactification, these branes have a series of images. Therefore, a signal sent to the bulk from the TeV brane has to come back to the original brane. The TeV brane is massive and curves the spacetime in such a way to allow for the bulk signal to reach a distant point on the TeV brane upon returning faster than the signal which propagates along the brane. Basically, a signal propagating along the brane is redshifted more (e.g. suffers a greater gravitational time delay) than the bulk signal, because of the presence of matter on the brane. While the signal never overtakes light in its own locality, it still allows for superluminal communication between two distant points on the brane  (if the signal is sent through the bulk). 

Unlike previous examples, where a specific shape of the brane was tailored in order to produce shortcuts, or a moving brane was invoked to break the Lorentz symmetry, our example is practically a generic (and yet realistic) variant of Randall-Sundrum models which fatefully reproduces FRW equations of the brane.  

Note that in this setup causality is not violated. The bulk signal is always light-like. The effect we describe here is very similar to gravitational lensing. A lens can produce two (or more) images of an object that can arrive to an observer at different times due to the curvature of space. In this process causality is not violated.     

Since the signals sent through the bulk always come back to the original brane and generically travel faster than signals along the brane, this effect might be used to resolve the cosmological horizon problem. 

All the discussion here was based on propagation of massless signals, so practically all of the conclusions drawn here will remain the same for gravitational waves propagation. 
One of the striking observational signatures  would be arrival of the same gravitational wave signal at two different times, where the first signals arrives before its electromagnetic counterpart.       
In addition, echo-like signals \cite{Dong:2020odp,Abedi:2016hgu} should also be present. We used GW170104 gravitational wave event to impose a strong limit on the model with extra dimensions in question.  Therefore we expect the size of compact space to be either so small that the effects of extra dimension are washed out, or very big( $>0.05$Mpc).

At the end, we note that the coincident detections of the gravitational wave signal GW170817 and gamma ray burts GRB170817A \cite{LIGOScientific:2017zic}, which indicate that gravitational waves and photons have to propagate with the same speeds up to the very high accuracy of $O(10^{-16})$, do not affect our limit.  In the context of the model that we used here, it just means that the detected gravitational waves and GRB signal propagated along the brane.

\begin{acknowledgments}
D.C. Dai is supported by the National Science and Technology Council (under grant no. 111-2112-M-259-016-MY3).
D.S. is partially supported by the US National Science Foundation, under Grant No.  PHY-2014021.  
\end{acknowledgments}

\end{document}